\documentclass[pdftex,10pt,a4paper]{article} 
\usepackage{xcolor}
\usepackage{graphicx}
\usepackage{amsmath}
\usepackage{url}

\title{Poiseuille Flow of a Non-Local Non-Newtonian Fluid with Wall Slip:\\
A First Step in Modeling Cerebral Microaneurysms}

\author{Corina S. Drapaca\\ Pennsyvania State University, University Park, PA 16802, USA\\}

\begin{document}

\maketitle

\abstract{Cerebral aneurysms and microaneurysms are abnormal vascular dilatations with high risk of rupture. An aneurysmal rupture could cause permanent disability and even death. Finding and treating aneurysms before their rupture is very difficult since symptoms can be easily attributed mistakenly to other common brain diseases. Mathematical models could highlight possible mechanisms of aneurismal development and suggest specialized biomarkers for aneurysms. Existing mathematical models of intracranial aneurysms focus on mechanical interactions between blood flow and arteries. However, these models cannot apply to microaneurysms since the anatomy and physiology at the length scale of cerebral microcirculation are different. In this paper we propose a mechanism for the formation of microaneurysms that involves the chemo-mechanical coupling of blood and endothelial and neuroglial cells. We model the blood as a non-local non-Newtonian incompressible fluid and solve analytically the Poiseuille flow of such a fluid through an axi-symmetric circular rigid and impermeable pipe in the presence of wall slip. The spatial derivatives of the proposed generalization of the rate of deformation tensor are expressed using Caputo fractional derivatives. The wall slip is represented by the classic Navier law and a generalization of this law involving fractional derivatives. Numerical simulations suggest that hypertension could contribute to microaneurysmal formation.}
\\
{\bf Keywords:} Fractional Calculus; Non-Newtonian Fluid; Poiseuille Flow; Cerebral Aneurysms; Microaneurysms.


\section{Introduction}

Cerebral aneurysms are abnormal swellings of the vasculature with high rupture risk. Although most cerebral aneurysms develop near bifurcations of the large arteries \cite{lie}, microaneurysms can also form on arterioles located either in the retinas of patients with diabetes \cite{agu03,new15} or deeper inside the brain tissue\\ 
\cite{ros77,ros08}. Aneurysmal ruptures could cause life threatening intracranial hemorrhages or, in the case of diabetic retinopathy, could lead to blindness.    
With the exception of retinal microaneurysms, finding cerebral aneurysms before they rupture is very hard because either there are no symptoms or the symptoms can be linked to other more common brain diseases. Effective biomarkers for cerebral aneurysms do not exist yet \cite{hus}. Mathematical models incorporating experimental and clinical observations of aneurysms can help us better understand aneurysmal pathophysiology, assess therapeutical success, identify biomarkers for the risk evaluation of aneurysmal formation and rupture, and develop pharmacological remedies for the inhibition of aneurysmal growth \cite{liang}. \\

Existing models of intracranial aneurysms are macroscopic and study the roles played by blood hemodynamics, arterial wall biomechanics and geometry, blood-wall mechanical interactions, mechanobiology, and the intracranial environment in the onset, evolution and rupture of aneurysms. With few exceptions, the blood is considered an incompressible Newtonian fluid in laminar flow through an artery of idealized geometry and having impermeable thin walls modeled as either rigid or viscoelastic solids. Recently, computer simulations of cerebral hemodynamics involving patient-specific geometries reconstructed from medical images and computational fluid dynamics (CFD) software have been performed and some geometrical and hemodynamical parameters were proposed as biomarkers for aneurysmal rupture prediction \cite{kul,jou,meng,ceb,ou,xu,ngoepe,grinberg}. A CFD model using an incompressible non-Newtonian constitutive law for the blood flowing with no-slip through a rigid-walled vessel was given in \cite{ceb}, while CFD models of incompressible Newtonian blood circulating with no-slip through vessels with deformable walls were presented in \cite{ou,xu}. The role played by blood clotting in aneurysmal growth and rupture was studied in \cite{grinberg} using a complex atomistic-continuum model of hemodynamics, and \cite{ngoepe} using mixture theory and diffusion-reaction equations for some blood proteins responsible for clotting formation and growth. A multi-particle collison dynamics approach was used in \cite{katrin} to simulate blood flow through an axi-symmetric circular cylinder with slip, and the mechanical response of a deformable arterial wall to mechanical forces exerted on it by the exterior surrounding structures was studied in \cite{hodis1,hodis2}. However, the mechanical parameters of these models lack predictive robustness and thus they cannot be used as biomarkers for aneurysms. In addition, the models require knowledge of numerous parameters which are not only difficult if not impossible to measure in vivo but also their relevance and established physical meanings might be obscured by the complex structure and dynamics possessed by this living system. This severely restricts the applicability of these models in clinical practice at this time. Lastly, the above mentioned mathematical models cannot be used in studying microaneurysms since different brain structures and chemo-mechanical interactions are relevant at this length scale. However, mathematical models for microaneurysms do not appear to exist yet although with the increasing prevalence of diabetes throughout the world \cite{bha} such models could be critical not only for understading the underlying mechanisms of retinal microaneurysms and diabetic retinopathy but also in developing effective treatments. \\

\begin{figure}[htbp]
\centering
\centerline{\includegraphics[width=10cm, scale=0.5]{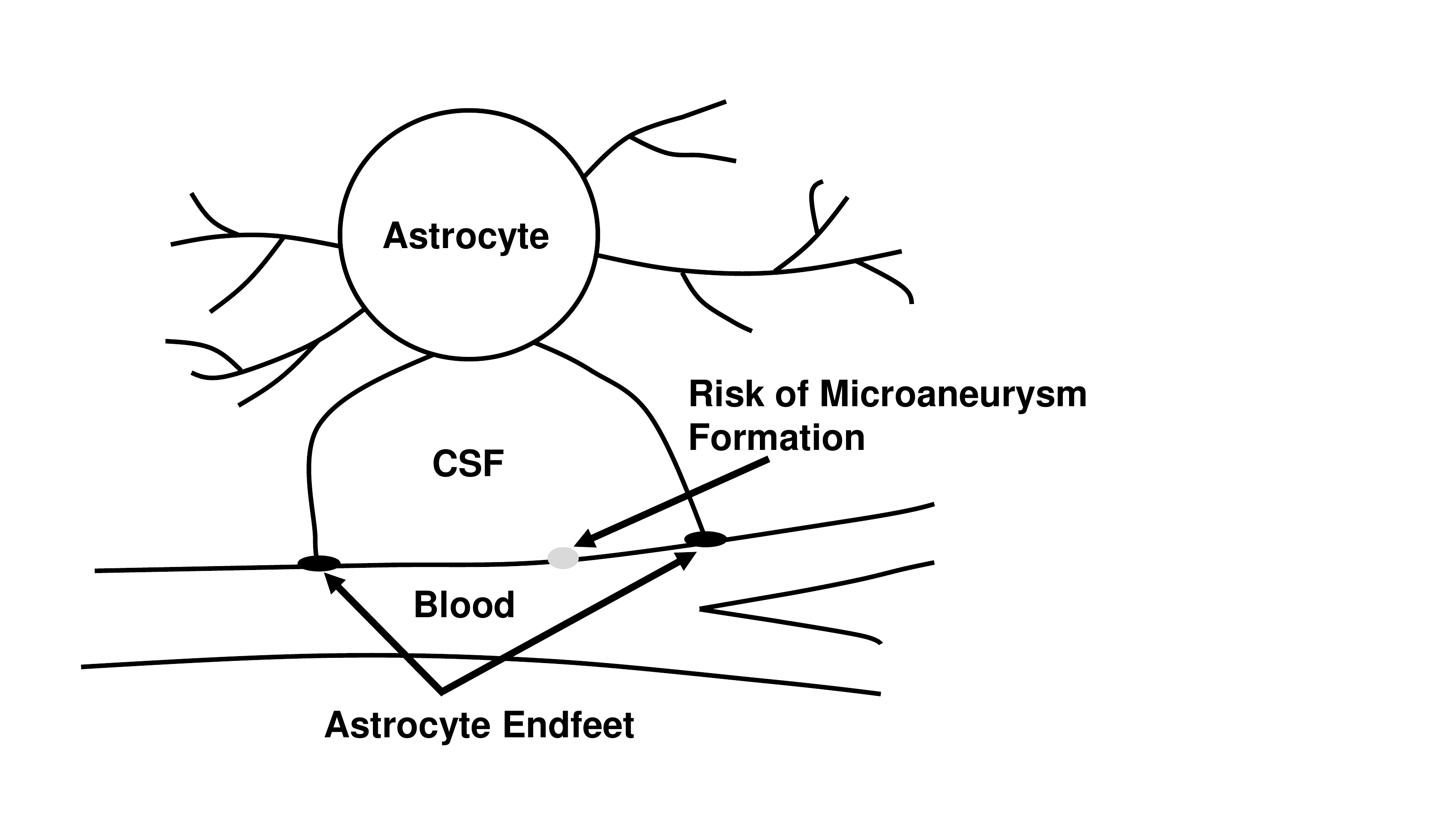}}
\caption{Schematic of the chemo-mechanical coupling among blood, cerebral arterioles and astrocytes. Slip at the blood-arteriole's wall interface could trigger an excessive vasodilation of the wall that is controlled by the astrocyte endfeet. This process could further lead to the formation of a microaneurysm. Brain cells and vasculature are immersed in cerebrospinal fluid (CSF), a water-like fluid produced by the brain which facilitates the intracellular and extracellular transport of molecules and ions within the brain.}
\label{fig1}
\end{figure}

 In this paper we propose the following mechanism for the formation of microaneurysms. A chemo-mechanical entanglement of blood and endothelial cells at the blood - vascular wall interface that might be caused by a chemical imbalance and/or the forward and backward moving waves reflected from a bifurcation site \cite{zamir} will cause slip of the blood at the wall. The slip will produce deviations from the expected shear stress level at the wall which will make the red blood cells release the energy-making enzyme called ATP \cite{fish} that will activate a chemical response from the endothelial cells of the wall which will lastly trigger neuronal-induced release of vasodilators from astrocytes whose endfeets continuously probe and control the amount of mechanical forces on the arterial wall and facilitate complex mechanotrasduction processes \cite{iade,met,mac,gor,pet}. According to \cite{mac,gor}, the two astrocyte endfeet use ${\rm Ca}^{2+}$-mediated chemical signals to control vasoconstriction (one endfoot) and vasodilation (the other endfoot). We hypothesize that slip could cause an excessive localized vasodilation which could lead to the formation of a microaneurysm. If some vasoconstriction is still present during this abnormal vasodilation, then the microaneurysm might evolve without rupture. However, if the vasoconstriction is missing, we conjecture that there is a high risk of microaneurysmal rupture. If this mechanism of microaneurysmal formation, growth and rupture is possible in practice then biomarkers for microaneurysms could most likely be found among chemical species located outside an arteriole at risk of microaneurysmal development. In figure \ref{fig1} we show a schematic of the structures that are involved in the formation of microaneurysms.      
 
 A mathematical model capable of predicting the formation, growth and risk of rupture of a microaneurysm needs to incorporate the well-known non-Newtonian behavior of blood when flowing through smaller vessels \cite{fung,fish,tom}, the deformability of the vascular wall \cite{mons}, and relevant mechanotransduction processes taking place among red blood cells, endothelial cells and astrocytes \cite{mac,gor}. The first step in building such a model is presented in this paper. Here we study the fully developed steady flow of blood through an arteriole due to an externally imposed pressure gradient and in the absence of body forces under the following assumptions: 1). blood is an incompressible non-Newtonian fluid with a non-local stress-deformation rate relationship, 2). the arteriole is a horizontal axi-symmetric circular pipe with rigid and impermeable walls, and 3) slip at the blood-vascular wall interface. In the constitutive equation of blood a generalized rate of deformation tensor is defined using non-local fractional order Caputo spatial derivatives of order $\alpha,\;m-1<\alpha<m,\, m=1,2,3...$. This is an extension of our model in \cite{dra}. The physical parameter $\alpha$ is seen as a measure of the long range interactions among fluid's particles during flow and it could vary with time, location, temperature, pressure, shear rate, and/or concentration of particles. In this paper we assume that $\alpha$ is constant. As is customary in fluid mechanics, the Cauchy stress tensor is a linear function of this rate of deformation tensor where the corresponding coefficient of proportionality $\mu$ that relates these two tensors is a physical parameter that depends on a characteristic length scale of the problem and the fractional order $\alpha$. In this paper we consider $\mu$ to be constant. We assume further that the flow has no slip at the wall until the shear stress at the wall reaches a critical value after which slip happens. The slip at the blood-vascular wall interface is modeled using the classic Navier slip condition and a new non-local generalization of Navier slip condition that involves a fractional order Caputo derivative. The parameter $\alpha$ in this generalized slip condition models the non-local entanglement and aggregation of the blood and endothelial cells. Thus we use fractional order derivatives in space to link cells from different structures. For now the fractional derivatives model the macroscopic behavior of aggregated blood cells and the macroscopic slip of entangled blood and endothelial cells. In our future work we plan to use the same approach involving spatial fractional order derivatives to account for the neuroglial effects on the arterioles, as well.\\    

Since at this first modeling step we do not consider the role played by the neuroglial dynamics in the vasodilation of arterioles, we show numerical simulations for arterioles and arteries for two characteristic slip lengths and various values of $\alpha \in (0,2)$. The results show that the velocity profiles and volumetric flow rates differ for the two slip lengths and for two values of $\alpha$ chosen such that one is less than 1 and the other is greater than 1. The velocity profiles and flow rates corresponding to the two slip conditions are relatively similar for the same slip length and value of $\alpha$, except for the case when the radius of the pipe is small (arteriole) and the slip length is big. In this case the velocity at the wall and the volumetric flow rate increase dramatically with a small increase in the pressure gradient. This suggests that if the blood pressure is consistently above a critical shear stress level responsible for slip which may happen for instance in the case of hypertension (caused by diabetes or not) then the produced blood flow rates will be persistently high. This repeatedly high flow rate will not only cause (mainly mechanically-driven) vasodilation but also activate, through mechanotransduction, neuroglial processes involved in the release of vasodilators that will further increase vasodilation which ultimately could lead to the formation of microaneurysms. Our findings are in agreement with some clinical observations \cite{ros77,ros08}. We emphasize that, to the best of our knowledge, this is the first time that such a mechanism for microaneurysms has been proposed in the literature. Its experimental and/or clinical validation will be the aim of future work.\\
    
In conclusion, the contributions of this paper are as follows: 
\begin{enumerate}
\item[1).] a mechanism for the formation, growth and rupture of micro-\\
aneurysms that involve the coupled dynamics of blood, arterioles and neuroglia was proposed, 
\item[2).] a non-local constitutive law involving spatial fractional order Caputo derivatives and only two physical parameters ($\alpha$ and $\mu$) was proposed to describe the non-Newtonian behavior of blood,
\item[3).] a non-local slip condition at the blood-vascular wall interface was introduced,
\item[4).] analytic solutions to a Poiseuille flow through an axi-symmetric circular rigid and impermeable pipe with wall slip were proposed, and
\item[5).] numerical simulations indicated that hypertension might contribute to the formation of microaneurysms which agrees with some clinical studies.
\end{enumerate}
From a physical point of view, the fractional order $\alpha$ is a measure of the non-local interactions of fluid particles during flow and could also control their deformability and aggregation patterns. The other physical parameter of the model, $\mu$, reduces to the apparent viscosity of the fluid when $\alpha=1$.\\

The structure of the paper is as follows. Our mathematical model and corresponding analytic solutions are presented in section 2. In section 3 we show and discuss some numerical simulations for blood flowing through an artery and an arteriole for various values of $\alpha$ and two characteristic slip lengths. The paper ends with a section of conclusions and future work.     

\section{Mathematical Model}

\subsection{Non-local Kinematics}

The following definition is a straightforward generalization of definition 6 in \cite{dra}:\\

{\bf Definition 1:} Let $\Omega$ and $\Omega_{t},\,t>0$ be two open subsets of ${\bf R}^{3}$. Let $\pmb{\alpha}$ be a 3x3 matrix whose elements $\alpha_{Ii}(t),\,I,i=1,2,3$ are continuous functions of $t > 0$ such that $\forall t>0,\,I,i=1,2,3$ either $-\infty<\alpha_{Ii}(t)<0$ or $m-1<\alpha_{Ii}(t)\leq m,\,m=1,2,3...$, and let $\pmb{\chi}(\cdot;\,t,\,\pmb{\alpha}(t)):\Omega \rightarrow \pmb{\chi}(\Omega;\,t,\,\pmb{\alpha}(t))\equiv\Omega_{t,\pmb{\alpha}(t)}$ be a family of functions in $L^{1}(\Omega)$. The {\it deformation of order} $\pmb{\alpha}(t)$ of a body occupying region $\Omega$  at $t=0$ and region $\Omega_{t,\,\pmb{\alpha}(t)}$ at time $t>0$ is determined by the position $\bf{x}$ of the material points in space as a function of the reference position ${\bf X}$ at $t=0$, time $t>0$ and $\pmb{\alpha}(t)$ which is given by:
\begin{enumerate}
\item[$\bullet$] if $\alpha_{Ii}(t)=m,\,\forall t>0,\;I,i=1,2,3,\,m=1,2,3,...$ and $\pmb{\chi}(\cdot;\,t,\,\pmb{\alpha}(t))$ is in $\mathcal{C}^{m+1}(\Omega)$:
\begin{equation}
x^{i}=\displaystyle \frac{\partial^{m-1}\chi^{i}({\bf X},t)}{\partial (X^{I})^{m-1}}
\label{def1_eq1}
\end{equation}
\item[$\bullet$] if $-\infty<\alpha_{Ii}(t)<0,\,\forall t>0,\,I,i=1,2,3$
\begin{align}
x^{i}=&\displaystyle \frac{1}{\Gamma(-\alpha_{1i}(t))\Gamma(-\alpha_{2i}(t))\Gamma(-\alpha_{3i}(t))} \nonumber\\
       \times & \iiint_H \frac{\chi^{i}({\bf Y},t) dY^{1}dY^{2}dY^{3}}{|X^{1}-Y^{1}|^{1+\alpha_{1i}(t)}
       |X^{2}-Y^{2}|^{1+\alpha_{2i}(t)}|X^{3}-Y^{3}|^{1+\alpha_{3i}(t)}},
\label{def1_eq2}
\end{align}
\item[$\bullet$] if $m-1<\alpha_{Ii}(t) < m,\,\forall t>0,\; I,i=1,2,3,\; m=1,2,3....$:
\begin{align}
\hspace{-0.5cm} x^{i}=&\displaystyle \frac{1}{\Gamma(m-\alpha_{1i}(t))\Gamma(m-\alpha_{2i}(t))\Gamma(m-\alpha_{3i}(t))} \nonumber\\
       \times & \frac{\partial^{m-1}}{\partial (X^{I})^{m-1}} \iiint_H \frac{\chi^{i}({\bf Y},t) dY^{1}dY^{2}dY^{3}}{|X^{1}-Y^{1}|^{1-m+\alpha_{1i}(t)}
       |X^{2}-Y^{2}|^{1-m+\alpha_{2i}(t)}|X^{3}-Y^{3}|^{1-m+\alpha_{3i}(t)}},
\label{def1_eq3}
\end{align}
\end{enumerate}
In addition, we assume that $\pmb{\chi}$ is zero on the boundary and outside the region of integration $H$. In equations (\ref{def1_eq2})-(\ref{def1_eq3}) $\displaystyle \Gamma(s)=\int_{0}^{\infty} t^{s-1} \exp(-t)dt$ is the gamma function.\\

Equation (\ref{def1_eq1}) for $m=1$ is the mathematical representation of deformation given in the classical theory of continuum mechanics, while the case $-\infty< \alpha_{Ii}(t)\leq 1,\,\forall t>0,\,I,i=1,2,3$ was introduced in \cite{dra}. Equation (\ref{def1_eq3}) proposed here accounts for possible effects of geometry (such as curvature) of microstructure on deformation.\\

The reasons for using fractional order integro-differential operators in equations (\ref{def1_eq2}) and (\ref{def1_eq3}) are threefold. In general, it is very difficult to use the physics of a problem to create non-local operators especially for materials with complex microstructural components that act and interact on multiple length and time scales. Fractional order operators model well spatio-temporal uncertainties in dynamical systems \cite{west} which is desirable in modeling for instance chemo-mechanical behavior of living biological tissues. Lastly, using these operators is mathematically convenient because we can apply known concepts from fractional calculus. Under certain mathematical assumptions the operators used in (\ref{def1_eq2}) and (\ref{def1_eq3}) could perhaps be fit into the general framework for non-local calculus proposed in \cite{du}. \\

According to \cite{dra}, the region of integration $H=[a_1,\,b_1]\times [a_2,\,b_2]\times [a_3,\,b_3]\subset {\bf R}^{3}$ respresents the region of influence of ${\bf X}$ and contains all the material points involved in the deformation of ${\bf X}$ into ${\bf x}$. The limits $a_i,\,b_i,\,i=1,2,3$ can be functions of ${\bf X},\,t$ and $\pmb{\alpha}$. $H$ can be determined by the physics of interactions between particles, by the geometry of the domain occupied by the body under observation, or by curve fitting to experimental data. The multiple length scales introduced by $H$ vary during deformation and are related through power laws of orders contained in the parameter matrix $\pmb{\alpha}(t)$. The components of matrix $\pmb{\alpha}(t)$ can be regarded as measures of the dynamic deformation of a body with evolving microstructure. \\

{\bf Definition 2:} The {\it deformation gradient of order} $\pmb{\alpha}(t)$, with $t>0$ and $\alpha_{Ii}\in {\bf R},\,I,i=1,2,3$ is either:
 \begin{equation}
 {\bf F}_{\pmb{\alpha}(t)}=\displaystyle \left(\frac{\partial x^{k}}{\partial X^{K}} \right)_{K,k=1,2,3},\;
{\rm if}\; \alpha_{Ii}(t)>0,\,\forall t>0,\,I,i=1,2,3,
 \label{def2_eq1}
 \end{equation}
 or:
 \begin{equation}
 {\bf F}_{\pmb{\alpha}(t)}={\bf x},\;{\rm if}\; -\infty < \alpha_{Ii}(t)<0,\,\forall t>0, \,I,i=1,2,3.
 \label{def2_eq2}
 \end{equation}
 
Definition 2 is an adaptation of definition 8 in \cite{dra}. We notice that the deformation gradient of order $\pmb{\alpha}(t)$ given by formulas (\ref{def2_eq1}) and (\ref{def2_eq2}) resembles the definitions of the left-sided and right-sided Riemann-Liouville fractional derivatives of order $\alpha_{Ii}(t),\,I,i=1,2,3$. Using some properties of fractional order derivatives it can be shown that inversions can be defined only for the deformations (\ref{def1_eq1}) and (\ref{def1_eq3}). The very long spatial memory of each material point during a deformation (\ref{def1_eq2}) is lost and thus an inversion of this motion is not possible. \\

Other strain and strain rate tensors of order $\pmb{\alpha}(t)$ can be further obtained by combining definitions known from continuum mechanics and formulas (\ref{def2_eq1}) or (\ref{def2_eq2}). For instance, the {\it velocity gradient tensor of order} $\pmb{\alpha}(t)$ is given by:
\[
{\bf L}_{\pmb{\alpha}(t)}=\displaystyle \left( \frac{\partial v^{k}}{\partial x^{k}} \right)_{k=1,2,3}
\]
where ${\bf v}$ is the velocity field associated to the deformation given in definition 1. It follows that the {\it rate of deformation tensor of order} $\pmb{\alpha}(t)$ is ${\bf D}_{\pmb{\alpha}(t)}=\left( {\bf L}_{\pmb{\alpha(t)}} + {\bf L}_{\pmb{\alpha(t)}}^{T} \right)/2$.  In particular, if $H$ is independent of $t$ and $\pmb{\alpha}(t)$ and $\pmb{\alpha}$ is a constant matrix, then the velocity fields $\displaystyle {\bf v}=\frac{\partial {\bf x}}{\partial t}$ corresponding to the three cases in definition 1 are given by formulas (\ref{def1_eq1})-(\ref{def1_eq3}) where $\pmb{\chi}$ is replaced by $\displaystyle \frac{\partial \pmb{\chi}}{\partial t}$. In this paper we will work under these simplifying assumptions and use the rate of deformation tensor of order $\pmb{\alpha}$ to model the non-local non-Newtonian behavior of blood.

\subsection{Poiseuille Flow}

We present now the mathematical model of the three-dimensional fully developed steady laminar flow of an incompressible non-local non-Newtonian fluid through an horizontal circular pipe with rigid and impermeable walls. The flow is axi-symmetric and is driven by an externally imposed pressure gradient. For simplicity, body forces are neglected. In cylindrical coordinates $(r,\theta,z)$, the components of fluid's velocity are: 
\begin{equation}
v_{r}=0,\;v_{\theta}=0,\;v_{z}=w(r),
\label{vel}
\end{equation}
 and therefore the equation of continuity:
\begin{equation}
\displaystyle \frac{\partial v_{r}}{\partial r}+\frac{v_{r}}{r}+\frac{\partial v_{z}}{\partial z}=0,
\label{ec}
\end{equation}
 is identically satisfied. The equilibrium equations involving the components of $\pmb{\sigma}(r,z)$, the Cauchy stress tensor of the fluid, in cylindrical coordinates are:
 \begin{align}
 \displaystyle \frac{\partial \sigma_{rr}}{\partial r}+\frac{1}{r} \frac{\partial \sigma_{r \theta}}{\partial \theta}+\frac{\partial \sigma_{rz}}{\partial z}+\frac{\sigma_{rr}-\sigma_{\theta \theta}}{r}&=0,\nonumber\\
 \displaystyle \frac{\partial \sigma_{r \theta}}{\partial r}+\frac{1}{r}\frac{\partial \sigma_{\theta \theta}}{\partial \theta}+
 \frac{\partial \sigma_{\theta z}}{\partial z}+\frac{2}{r}\sigma_{r \theta}&=0,\nonumber\\
 \displaystyle \frac{\partial \sigma_{rz}}{\partial r}+\frac{1}{r}\frac{\partial \sigma_{\theta z}}{\partial \theta}+\frac{\partial \sigma_{zz}}{\partial z}+\frac{1}{r}\sigma_{rz}&=0.
 \label{ee}
 \end{align}
 The only non-zero components of $\pmb{\sigma}$ are given by the following constitutive formulas:
 \begin{align}
 \sigma_{rr}&=\sigma_{\theta \theta}=\sigma_{zz}=-p(r,z), \label{claw_p} \\
 \sigma_{rz}&=\sigma_{zr}=\displaystyle \mu \frac{1}{\Gamma(m-\alpha)}\frac{d^{m}}{dr^{m}}\int_{0}^{r} \frac{1}{(r-\tau)^{\alpha+1-m}}\nonumber\\
 &\times \left( w(\tau) - \sum_{k=0}^{m-1}\frac{\tau^{k}}{k!}\frac{d^{k}}{d\tau^{k}}w(0^+)\right)d\tau=\mu D_{r}^{\alpha}w(r) \label{claw_shear}
 \end{align}
where $p$ is the hydrostatic pressure of the fluid, and $\mu$ and $\alpha\in (m-1,m),\,m=1,2,3...$ are the two physical parameters of the model. In the constitutive formula (\ref{claw_shear}), the shear stress is assumed to be proportional to the shear rate of order $\alpha$ introduced earlier. Here the region of influece is $H=[0,r]$ where the lower limit represents the axis of symmetry of the pipe $r=0$. We assume that  $\displaystyle \frac{d^{m}w}{dr^{m}}$ is in $L^{1}([0,R])$, where $R$ is the constant radius of the pipe, and introduce an extra term under the integral in (\ref{claw_shear}) which is not present in definition 1 so that we can impose non-zero boundary conditions. In this case the integral in (\ref{claw_shear}) is equal to the left-sided Caputo fractional derivative of order $\alpha$ \cite{main} which we denote here by $D_{r}^{\alpha}w(r)$. By definition, the left-sided Caputo fractional derivative of order $\alpha$ is \cite{main}:
 \begin{equation}
 D_{r}^{\alpha}w(r)=\begin{cases} J^{m-\alpha}\frac{d^{m}w(r)}{dr^{m}}= \displaystyle \frac{1}{\Gamma(m-\alpha)}\int_{0}^{r} \frac{\frac{d^{m}w(\tau)}{d\tau^{m}}}{(r-\tau)^{\alpha+1-m}}d\tau, & m-1<\alpha < m; \\
 \frac{d^{m}w(r)}{dr^{m}}, & \alpha = m;
 \end{cases}
  \label{caputo}
 \end{equation} 
 where $m \in\{ 1,2,3....\}$. 
In formula (\ref{caputo}) the definition of the Riemann fractional integral operator of order $\alpha$ denoted by $J^{\alpha}$ was also given \cite{main}. \\  

For simplicity, we assume further that\footnote{We expect that for more complex flows other boundary conditions for $\frac{d^{k}w}{dr^{k}}(0^{+})$ could be specified.}:
\begin{equation} 
\frac{d^{k}}{dr^{k}}w(0^+)=0,\,k=1,2,...m-1
\label{zero}
\end{equation}

In particular, if $\alpha=1$  equation (\ref{claw_shear}) becomes the well-known constitutive law of an incompressible viscous Newtonian fluid where $\mu$ is fluid's apparent viscosity.

To get an intuition into the physical meaning of parameter $\mu$, we perform now a dimensional analysis for equation (\ref{claw_shear}). We denote by $M,\,L,\,$ and $T$ the characteristic mass, length, and time scales, respectively. Then, the physical units of the quantities in equation (\ref{claw_shear}) are:
\begin{equation}
\displaystyle \left[\sigma_{rz}\right]= \frac{M}{LT^{2}},\; [D_{r}^{\alpha}w]=\frac{1}{L^{\alpha-1}T}
\label{da}
\end{equation}
It follows then from (\ref{claw_shear}) that the physical unit of $\mu$ is:
\begin{equation}
[\mu]=\displaystyle \frac{M}{L^{2-\alpha}T}
\label{damu}
\end{equation}
Formula (\ref{damu}) suggests that $\mu$ is a parameter that could for instance be represented as:
\begin{equation}
\mu=\mu_{0}L^{\alpha-1}
\label{mu1}
\end{equation}
where $\mu_{0}$ has the physical unit of apparent viscosity: $\left[\mu_{0}\right]=\frac{M}{LT}$.\\
Another representation of $\mu$ could be:
\begin{equation}
\mu=\displaystyle J^{\alpha}_{x}\mu_{0}
\label{mu2}
\end{equation}
where $x$ is a preferred direction of fluid's particles aggregation during flow which may not coincide with the flow direction, and $J^{\alpha}_{x}$ is the Riemann fractional integral operator of order $\alpha$. \\

\begin{figure}[htbp]
\centerline{\includegraphics[clip, trim=1cm 8.5cm 1cm 9cm, width=\textwidth]{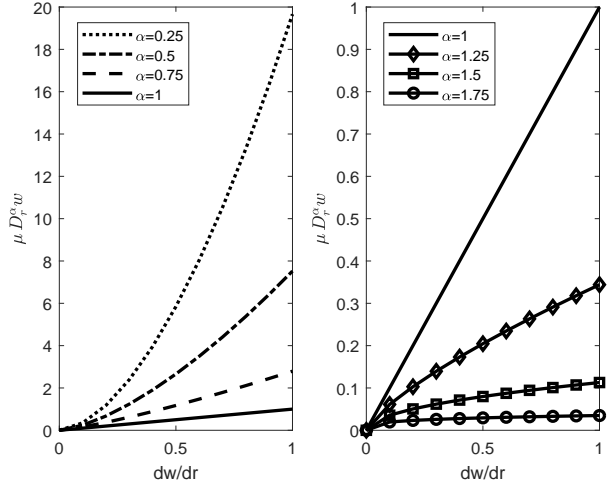}}
\caption{Shear stress versus classical shear rate for various values of $\alpha$: $0<\alpha\leq 1$ (left plot), and $1\leq \alpha<2$ (right plot).}
\label{fig2new}
\end{figure}

Let assume now that $\mu$ is given by (\ref{mu1}) and $\mu_{0}$ and $L$ are constant. Then, according to (\ref{claw_shear}), the shear stress varies linearly with the shear rate of fractional order $\alpha$, but not with respect to the classical shear rate unless of course $\alpha=1$. A simple calculation can tell us how the shear rate (\ref{claw_shear}) varries with the classical shear rate. If the classical shear rate $\frac{dw}{dr}$ is linear in the variable $r$, then $w=r^{2}/2$ and \cite{main}:
\begin{equation}
D_{r}^{\alpha}w=  \displaystyle \frac{r^{2-\alpha}}{\Gamma(3-\alpha)}
\label{example}
\end{equation}
In figure \ref{fig2new} we show the variations of the shear stress obtained by replacing (\ref{example}) into (\ref{claw_shear}) with the classical shear rate for $\mu_{0}=1,\,L=0.01$ (with generic physical units) and $\alpha \in \{
0.25, 0.5, 0.75, 1, 1.25, 1.5, 1.75\}$. The shear thinning observed experimentally for non-Newtonian fluids including the blood flow through small vessels is obtained for $1<\alpha<2$ (right plot of figure \ref{fig2new}). This suggests that the fractional order $\alpha$ is a parameter that provides an intinsic linkage between flow and the continuous evolution and rearrangement of fluid's microstructure during flow. When $\alpha=1$ the information about this coupling is lost and other ad-hoc assumptions are then to be made about the dependency of the apparent viscosity on the shear rate so that shear-thinning can be predicted by the constitutive law. Constitutive laws of non-Newtonian fluids that are obtained for $\alpha=1$ in the manner described here involve usually more physical parameters that need to be found experimentally. On the other hand, our constitutive model (\ref{claw_shear}) has only two physical parameters, $\mu$ and $\alpha$, which is a desirable feature especially when trying to observe blood flow in a living body in vivo with minimally invasive procedures.\\ 

In this paper, we assume that $\mu$ is constant\footnote{We notice that the analytical solutions presented further remain valid if $\mu$ is given by formula (\ref{mu1}) in the slip region where the characteristic length is assumed to be the slip length.}. Then, by replacing equations (\ref{claw_p}) and (\ref{claw_shear}) into the system of equations (\ref{ee}) we obtain:
\begin{eqnarray}
\displaystyle \frac{\partial p}{\partial r}=0, \label{pr} \\
\displaystyle \mu \frac{1}{r} \frac{\partial}{\partial r} \left( r\,D_{r}^{\alpha}w(r) \right)=\frac{\partial p}{\partial z} \label{wr}
\end{eqnarray} 
Equations (\ref{pr}) and (\ref{wr}) suggest that $\displaystyle \frac{dp}{dz}=C<0$ is a given constant. Integrating once equation (\ref{wr}) gives the following expression for the shear rate:
\begin{equation}
\displaystyle D^{\alpha}_{r}w(r)=\frac{C}{2\mu}r+\frac{c_{1}}{r}
\label{shear_r}
\end{equation}
where $c_{1}$ is a constant of integration. One boundary condition requires the fractional shear rate (\ref{shear_r}) to be bounded at $r=0$. This implies $c_{1}=0$ and thus:
\begin{equation}
\displaystyle D^{\alpha}_{r}w(r)=\frac{C}{2\mu}r
\label{shear_r1}
\end{equation}
We apply the operator $J^{\alpha}$ to (\ref{shear_r1}) and use the following properties of fractional integrals and derivatives \cite{main}:
\begin{eqnarray*}
\displaystyle J^{\alpha}D_{r}^{\alpha}w(r)=w(r)-\sum_{k=0}^{m-1}\frac{d^{k}w}{dr^{k}}(0^{+})\frac{r^{k}}{k!},\;m-1<\alpha<m\\
\displaystyle J^{\alpha}r^{\gamma}=\frac{\Gamma(\gamma+1)}{\Gamma(\gamma+1+\alpha)}r^{\gamma+\alpha},\; r>0,\,\alpha\geq 0,\;\gamma>-1
\end{eqnarray*}
to get:
\begin{equation}
\displaystyle w(r)=\frac{C}{2\mu \alpha (\alpha+1)\Gamma(\alpha)}r^{\alpha+1}+\sum_{k=0}^{m-1}\frac{d^{k}w}{dr^{k}}(0^{+})\frac{r^{k}}{k!}.
\label{sol_w}
\end{equation}
To obtain a physically consistent velocity, the second term of (\ref{sol_w}) will be found from the boundary condition at the wall $r=R$. We denote by $\tau_{w}=\sigma_{rz}|_{r=R}$ the shear stress at the wall and by $\tau_{c}>0$ the critical shear stress at which slip starts. Inspired by \cite{fer1,fer2,kao}, we impose the following boundary condition:
\begin{eqnarray}
w(R)=\begin{cases}
0,\; |\tau_{w}| \leq \tau_{c}\\
v_{slip}, \; |\tau_{w}|> \tau_{c}
\end{cases}
\label{slip_bc}
\end{eqnarray}
where the prescribed slip velocity $v_{slip}$ will be specified shortly.\\

By combining (\ref{sol_w}), (\ref{slip_bc}) and (\ref{zero}) we find:
\begin{eqnarray}
 \displaystyle w(r)=\begin{cases}
 \frac{C}{2\mu \alpha (\alpha+1) \Gamma(\alpha)} \left( r^{\alpha+1}-R^{\alpha+1}\right) ,\; |\tau_{w}| \leq \tau_{c}\\
 \frac{C}{2\mu \alpha (\alpha+1) \Gamma(\alpha)} \left( r^{\alpha+1}-R^{\alpha+1}\right) +v_{slip},\; |\tau_{w}| > \tau_{c}
 \end{cases}
 \label{speed}
 \end{eqnarray}
 
 \subsection{Two Slip Conditions}
 
 We provide now two expressions for the slip velocity $v_{slip}$. The first expression is the classic Navier slip condition \cite{fer1,fer2,kao}:
 \begin{equation}
 v_{slip}=-l \displaystyle \frac{dw}{dr}|_{r=R}
 \label{slip1}
 \end{equation}
 where $l$ is a characteristic slip length and the negative sign suggests the friction between the fluid motion and the wall. \\
 The second expression that we propose here is a novel generalization of the Navier slip condition:
 \begin{equation}
v_{slip}=  -l^{\alpha} D_{r}^{\alpha}w|_{r=R}
 \label{slip2}
 \end{equation}
 where the power law $l^{\alpha}$ is needed for dimensional consistency. We notice that if $\alpha=1$ then formula (\ref{slip2}) reduces to (\ref{slip1}).  The slip condition (\ref{slip2}) models the non-local entanglement and aggregation of the wall and fluid particles.\\
 
By substituting formula (\ref{speed}) into expression (\ref{slip1}) and respectively formula (\ref{shear_r1}) into expression (\ref{slip2}) the following formulas for $v_{slip}$ are obtained:
\begin{equation}
\displaystyle v_{slip}=-l \frac{C R^{\alpha}}{2\mu \alpha \Gamma(\alpha)}
\label{vslip1}
\end{equation}
and respectively:
\begin{equation}
\displaystyle v_{slip} = - l^{\alpha} \frac{C R}{2\mu}.
\label{vslip2}
\end{equation}
From formulas (\ref{claw_shear}) and (\ref{shear_r1}) we have:
\[
\tau_{w}=\frac{C R}{2}
\]
and thus we define $\tau_{c}=\frac{C^{\star}R}{2}$ with $C^{\star}>0$ a constant. We now replace the expressions for $\tau_{w}$ and $\tau_{c}$, and formulas (\ref{vslip1}), (\ref{vslip2}) into (\ref{speed}) and obtain the following formulas for the fluid velocity:
\begin{eqnarray}
 \displaystyle w(r)=\begin{cases}
 \frac{|C|R^{\alpha+1}}{2\mu \alpha (\alpha+1) \Gamma(\alpha)} \left( 1 - \left( \frac{r}{R}\right)^{\alpha+1}\right) ,\; |C| \leq C^{\star}\\
 \\
  \frac{|C|R^{\alpha+1}}{2\mu \alpha (\alpha+1) \Gamma(\alpha)} \left( 1 - \left( \frac{r}{R}\right)^{\alpha+1}+(\alpha+1) \frac{l}{R}\right) ,\; |C| > C^{\star}
 \end{cases}
 \label{speed_NSC}
 \end{eqnarray}
 corresponding to the Navier slip condition, and:
 \begin{eqnarray}
 \displaystyle w(r)=\begin{cases}
 \frac{|C|R^{\alpha+1}}{2\mu \alpha (\alpha+1) \Gamma(\alpha)} \left( 1 - \left( \frac{r}{R}\right)^{\alpha+1}\right) ,\; |C| \leq C^{\star}\\
 \\
  \frac{|C|R^{\alpha+1}}{2\mu \alpha (\alpha+1) \Gamma(\alpha)} \left( 1 - \left( \frac{r}{R}\right)^{\alpha+1}\right)+\frac{|C|R^{\alpha+1}}{2\mu}\left( \frac{l}{R} \right)^{\alpha} ,\; |C| > C^{\star}
 \end{cases}
 \label{speed_GNSC}
 \end{eqnarray}
corresponding to the generalized Navier slip condition.\\

Lastly, we calculate the volumetric flow rate which, in our case, is given by:
\[
\displaystyle Q = \int_{0}^{R} 2\pi\,r\,w(r) dr
\]
By using formula (\ref{speed_NSC}) we get:
\begin{eqnarray}
 \displaystyle Q =\begin{cases}
 \frac{\pi\,|C| R^{\alpha+3}}{2\mu \alpha (\alpha+3) \Gamma(\alpha)}  ,\; |C| \leq C^{\star}\\
 \\
  \frac{\pi\,|C|R^{\alpha+3}}{2\mu \alpha (\alpha+3) \Gamma(\alpha)} +\frac{\pi\,|C|\,l\,R^{\alpha+2}}{2\mu \alpha \Gamma(\alpha)} ,\; |C| > C^{\star}
 \end{cases}
 \label{Q_NSC}
 \end{eqnarray}
and by using formula (\ref{speed_GNSC}):
\begin{eqnarray}
 \displaystyle Q =\begin{cases}
 \frac{\pi\,|C| R^{\alpha+3}}{2\mu \alpha (\alpha+3) \Gamma(\alpha)}  ,\; |C| \leq C^{\star}\\
 \\
  \frac{\pi\,|C|R^{\alpha+3}}{2\mu \alpha (\alpha+3) \Gamma(\alpha)} +\frac{\pi\,|C|\,l^{\alpha}\,R^{3}}{2\mu} ,\; |C| > C^{\star}
 \end{cases}
 \label{Q_GNSC}
 \end{eqnarray}

\subsection{Non-Dimensionalization}

We introduce the following non-dimensional quantities:
\begin{equation}
\displaystyle \tilde{r}=\frac{r}{R},\;\tilde{C}=\frac{|C|}{C^{\star}}=\frac{|C|R}{2\tau_{c}},\; 
\tilde{w} = w\frac{\mu}{\tau_{c}R^{\alpha}},\;
\tilde{Q}=Q\frac{\mu}{\pi \,\tau_{c}R^{\alpha+2}}
\label{no_dim}
\end{equation}
The above formulas were adapted to our case from \cite{kao}. The quantities $\tilde{w}$ and $\tilde{Q}$ given by (\ref{no_dim}) are dimensionless because of formula (\ref{damu}). \\

Then the dimensionless representations of formulas (\ref{speed_NSC}) and (\ref{speed_GNSC}) are:
\begin{eqnarray}
 \displaystyle \tilde{w}(\tilde{r})=\begin{cases}
 \frac{\tilde{C}}{\alpha (\alpha+1) \Gamma(\alpha)} \left( 1 - \tilde{r}^{\alpha+1}\right) ,\;  \tilde{C}\leq 1\\
 \\
  \frac{\tilde{C}}{\alpha (\alpha+1) \Gamma(\alpha)} \left( 1 - \tilde{r}^{\alpha+1}+(\alpha+1) \frac{l}{R}\right) ,\; \tilde{C} > 1
 \end{cases}
 \label{nodim_speed_NSC}
 \end{eqnarray}
 and respectively:
 \begin{eqnarray}
 \displaystyle \tilde{w}(\tilde{r})=\begin{cases}
 \frac{\tilde{C}}{\alpha (\alpha+1) \Gamma(\alpha)} \left( 1 - \tilde{r}^{\alpha+1}\right) ,\; \tilde{C} \leq 1\\
 \\
  \frac{\tilde{C}}{\alpha (\alpha+1) \Gamma(\alpha)} \left( 1 - \tilde{r}^{\alpha+1}\right)+\tilde{C}\left( \frac{l}{R} \right)^{\alpha} ,\; \tilde{C} > 1
 \end{cases}
 \label{nodim_speed_GNSC}
 \end{eqnarray}

The dimensionless representations of formulas (\ref{Q_NSC}) and (\ref{Q_GNSC}) are:
\begin{eqnarray}
 \displaystyle \tilde{Q} =\begin{cases}
 \frac{\tilde{C}}{\alpha (\alpha+3) \Gamma(\alpha)} ,\; \tilde{C} \leq 1\\
 \\
  \frac{\tilde{C}}{\alpha (\alpha+3) \Gamma(\alpha)} +\frac{\tilde{C}}{\alpha \Gamma(\alpha)}\frac{l}{R} ,\; \tilde{C} > 1
 \end{cases}
 \label{nodim_Q_NSC}
 \end{eqnarray}
and respectively:
\begin{eqnarray}
 \displaystyle \tilde{Q} =\begin{cases}
 \frac{\tilde{C}}{\alpha (\alpha+3) \Gamma(\alpha)}  ,\; \tilde{C} \leq 1\\
 \\
  \frac{\tilde{C}}{\alpha (\alpha+3) \Gamma(\alpha)} +\tilde{C} \left( \frac{l}{R} \right)^{\alpha} ,\; \tilde{C} > 1
 \end{cases}
 \label{nodim_Q_GNSC}
 \end{eqnarray}
Formulas (\ref{nodim_speed_NSC})-(\ref{nodim_Q_GNSC}) are used in the results section.

\section{Results}

We present the effects of the pipe radius $R$, characteristic slip length $l$, and fractional order $\alpha$ on the fluid's velocity and volumetric flow rate. Since the application we are interested in is cerebral blood flow and in our mathematical model we did not incorporate yet possible neuroglial effects on the vasculature, we show results for flow through an artery and an arteriole. We used $R=0.25\,{\rm cm}$ for the radius of a cerebral artery \cite{jain}, and $R=0.001\,{\rm cm}$ for the radius of a cerebral arteriole which we estimated from the chosen arterial radius by applying Murray's law\footnote{Murray's law says that at a bifurcation of the blood vasculature the cube of the radius of a parent vessel equals the sum of the cubes of the radii of the two daughter vessels. In particular, if the bifurcation is symmetric then the radii of the two daughter vessels are equal.}  to a 24-level branching tree of symmetric bifurcations \cite{zamir}. We used two values for the characteristic slip length: 0.0002 cm, and 0.02 cm, which are approximately 10 times less and respectively 10 times more than the diameter of an endothelial cell (about 20 microns). The results for a characteristic slip length of $0.002\,{\rm cm}$ are similar to those for $l=0.0002\,{\rm cm}$ and therefore we will not present them. The fractional order $\alpha$ varies from 0.1 to 1.7, and we used the values of 0.5 and 1.7 to compare the flows corresponding to the classic Navier slip condition and its generalization proposed in this paper. \\

\begin{figure}[htbp]
\centerline{\includegraphics[clip, trim=1cm 8.5cm 1cm 9cm, width=\textwidth]{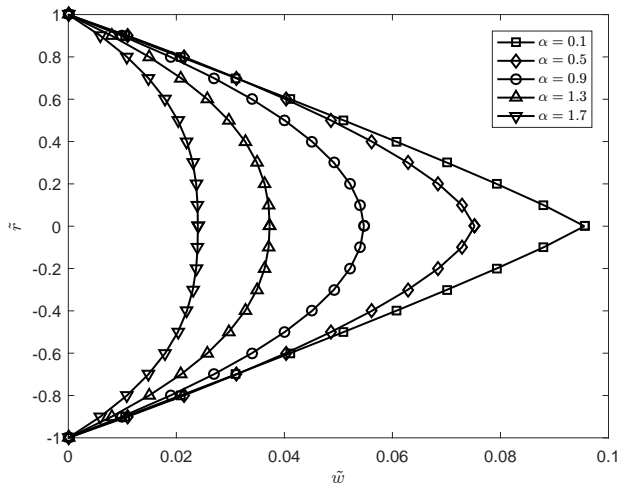}}
\caption{The non-dimensional velocity corresponding to the no-slip boundary condition calculated using either formula (\ref{nodim_speed_NSC}) or (\ref{nodim_speed_GNSC}) for $\tilde{C}=0.1$ and various values of $\alpha$.}
\label{fig2}
\end{figure}

In figure \ref{fig2} we show the dimensionless velocity of the fluid in no slip flow. We notice that the velocity profile for $\alpha=0.9$ is close to the classic Poiseuille flow of a Newtonian fluid, while the profile for $\alpha=1.7$ resembles blood's velocity observed experimentally \cite{san,zhong}.         

\begin{figure}[htbp]
\centerline{\includegraphics[clip, trim=1cm 8.5cm 1cm 9cm, width=\textwidth]{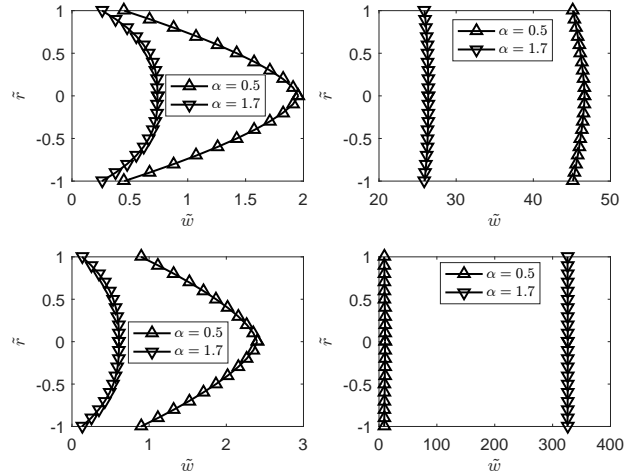}}
\caption{Non-dimensional velocities calculated using formulas (\ref{nodim_speed_NSC}) and (\ref{nodim_speed_GNSC}) for $\tilde{C}=2, \;R=0.001\, {\rm cm}$ and two values of $\alpha$: 0.5 and 1.7. The top row corresponds to the classic Navier slip condition, while the bottom row represents the speed for the generalized Navier slip condition. The left column is for $l=0.0002 \,{\rm cm}$, and the right column is for $l=0.02\,{\rm cm}$.}
\label{fig3}
\end{figure}

\begin{figure}[htbp]
\centerline{\includegraphics[clip, trim=1cm 8.5cm 1cm 9cm, width=\textwidth]{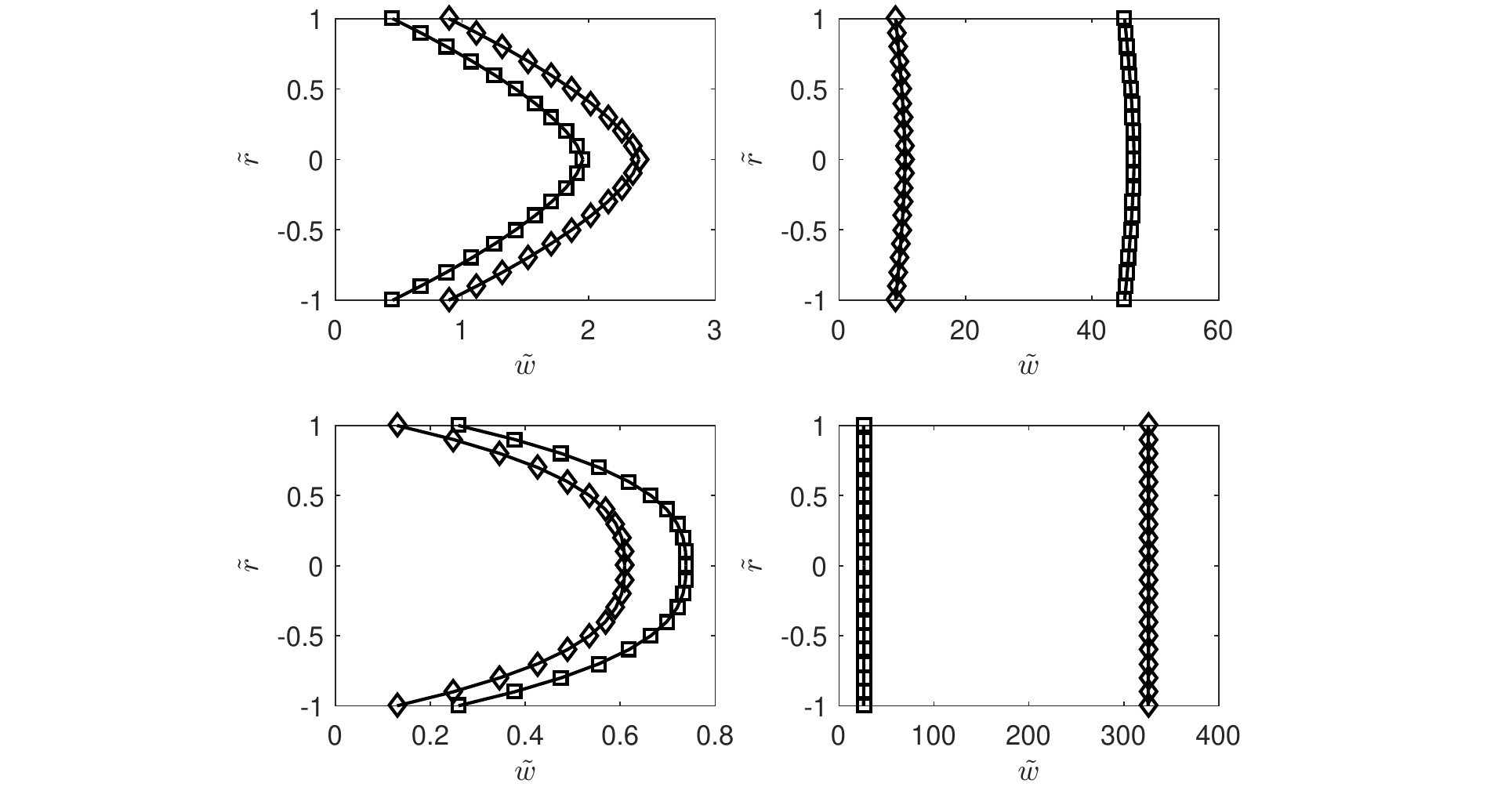}}
\caption{Non-dimensional velocities calculated using formulas (\ref{nodim_speed_NSC}) and (\ref{nodim_speed_GNSC}) for $\tilde{C}=2, \;R=0.001\, {\rm cm}$. The square symbol is used for the flow with classic Navier slip, and the diamond stands for the flow satisfying the generalized Navier slip condition. The top row corresponds to $\alpha=0,5$, while the bottom row corresponds to $\alpha=1.7$. The left column is for $l=0.0002 \,{\rm cm}$, and the right column is for $l=0.02\,{\rm cm}$.}
\label{fig4}
\end{figure}

\begin{figure}[htbp]
\centerline{\includegraphics[clip, trim=1cm 8.5cm 1cm 9cm, width=\textwidth]{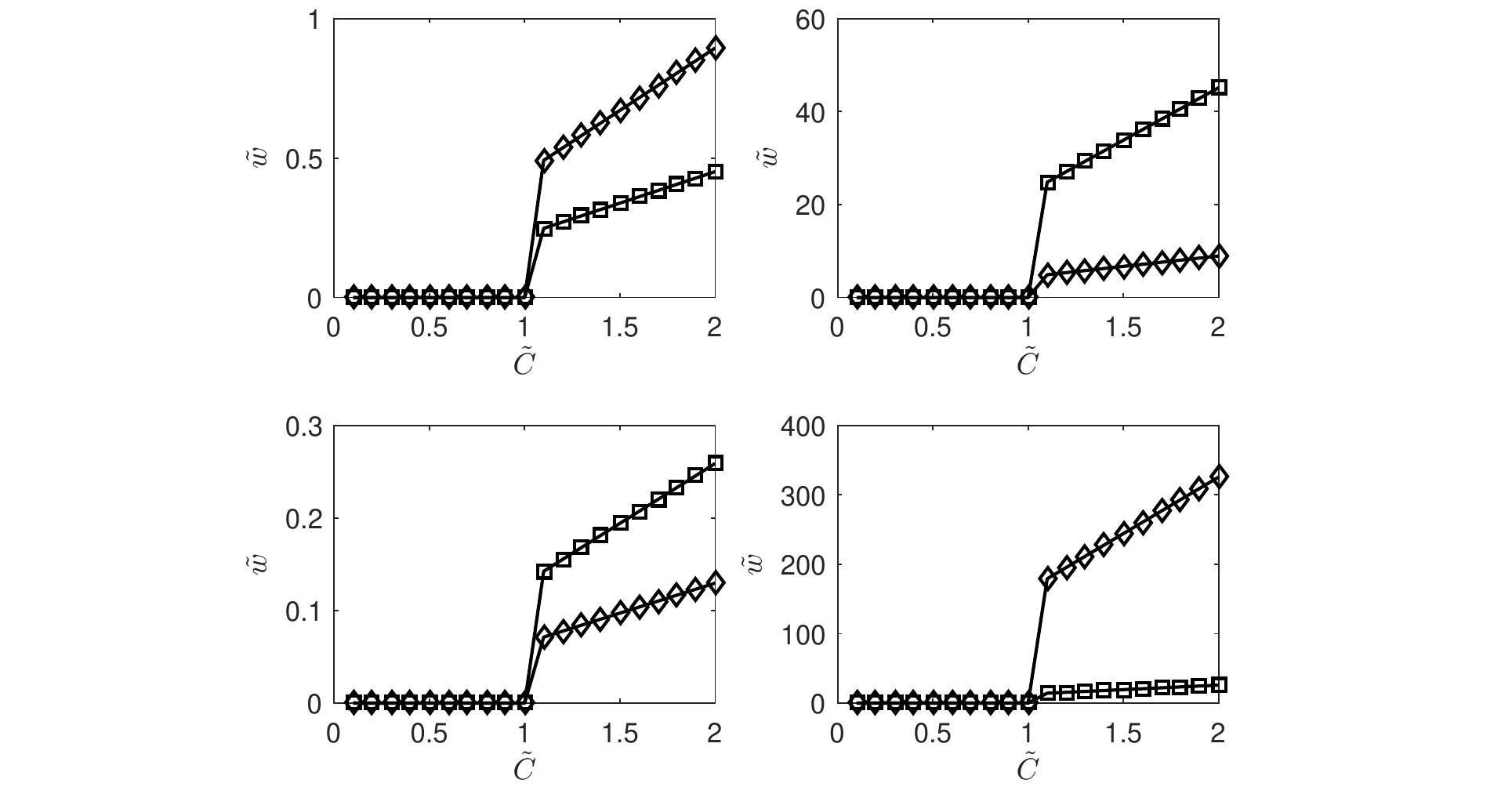}}
\caption{Non-dimensional velocities calculated at the wall using formulas (\ref{nodim_speed_NSC}) and (\ref{nodim_speed_GNSC}) for $R=0.001\, {\rm cm}$. The square symbol is used for the flow with classic Navier slip, and the diamond stands for the flow satisfying the generalized Navier slip condition. The top row corresponds to $\alpha=0,5$, while the bottom row corresponds to $\alpha=1.7$. The left column is for $l=0.0002 \,{\rm cm}$, and the right column is for $l=0.02\,{\rm cm}$.}
\label{fig5}
\end{figure}

\begin{figure}[htbp]
\centerline{\includegraphics[clip, trim=1cm 8.5cm 1cm 9cm, width=\textwidth]{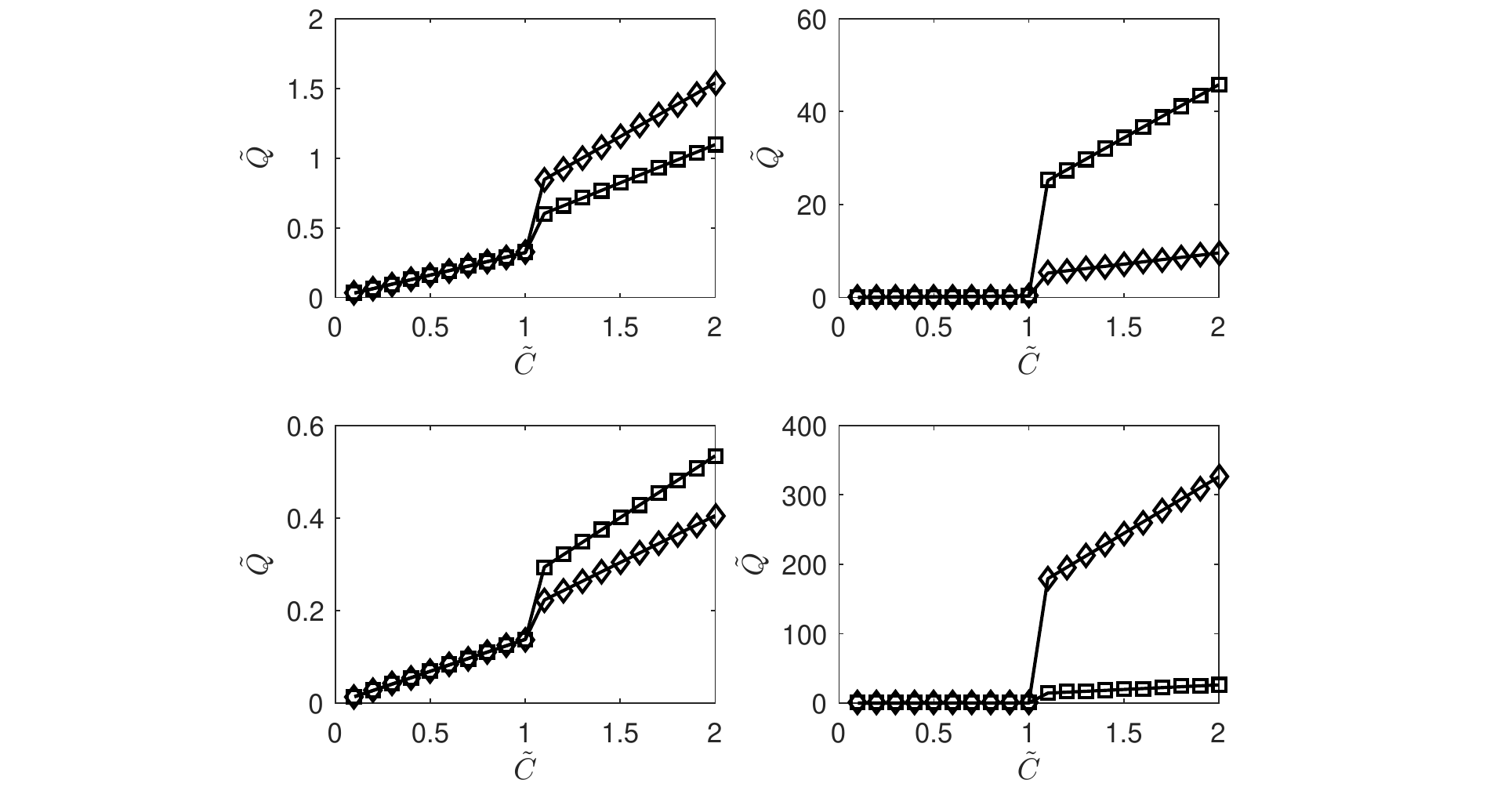}}
\caption{Non-dimensional volumetric flow rates calculated using formulas (\ref{nodim_Q_NSC}) and (\ref{nodim_Q_GNSC}) for $R=0.001\, {\rm cm}$. The square symbol is used for the flow with classic Navier slip, and the diamond stands for the flow satisfying the generalized Navier slip condition. The top row corresponds to $\alpha=0,5$, while the bottom row corresponds to $\alpha=1.7$. The left column is for $l=0.0002 \,{\rm cm}$, and the right column is for $l=0.02\,{\rm cm}$.}
\label{fig6}
\end{figure}

\begin{figure}[htbp]
\centerline{\includegraphics[clip, trim=1cm 8.5cm 1cm 9cm, width=\textwidth]{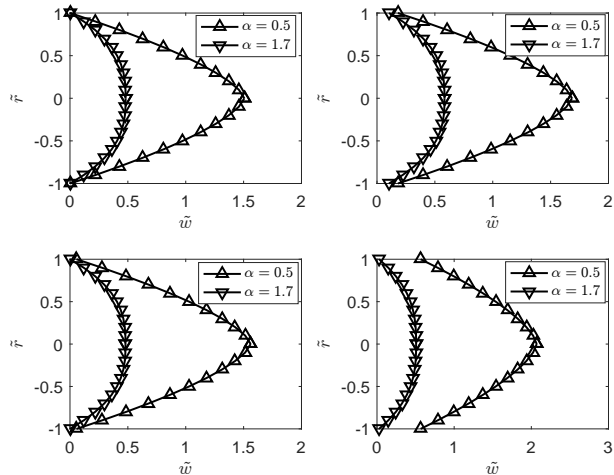}}
\caption{Non-dimensional velocities calculated using formulas (\ref{nodim_speed_NSC}) and (\ref{nodim_speed_GNSC}) for $\tilde{C}=2, \;R=0.25\, {\rm cm}$ and two values of $\alpha$: $\alpha=0.5$ and $\alpha=1.7$. The top row corresponds to the classic Navier slip condition, while the bottom row represents the speed for the generalized Navier slip condition. The left column is for $l=0.0002 \,{\rm cm}$, and the right column is for $l=0.02\,{\rm cm}$.}
\label{fig7}
\end{figure}

\begin{figure}[htbp]
\centerline{\includegraphics[clip, trim=1cm 8.5cm 1cm 9cm, width=\textwidth]{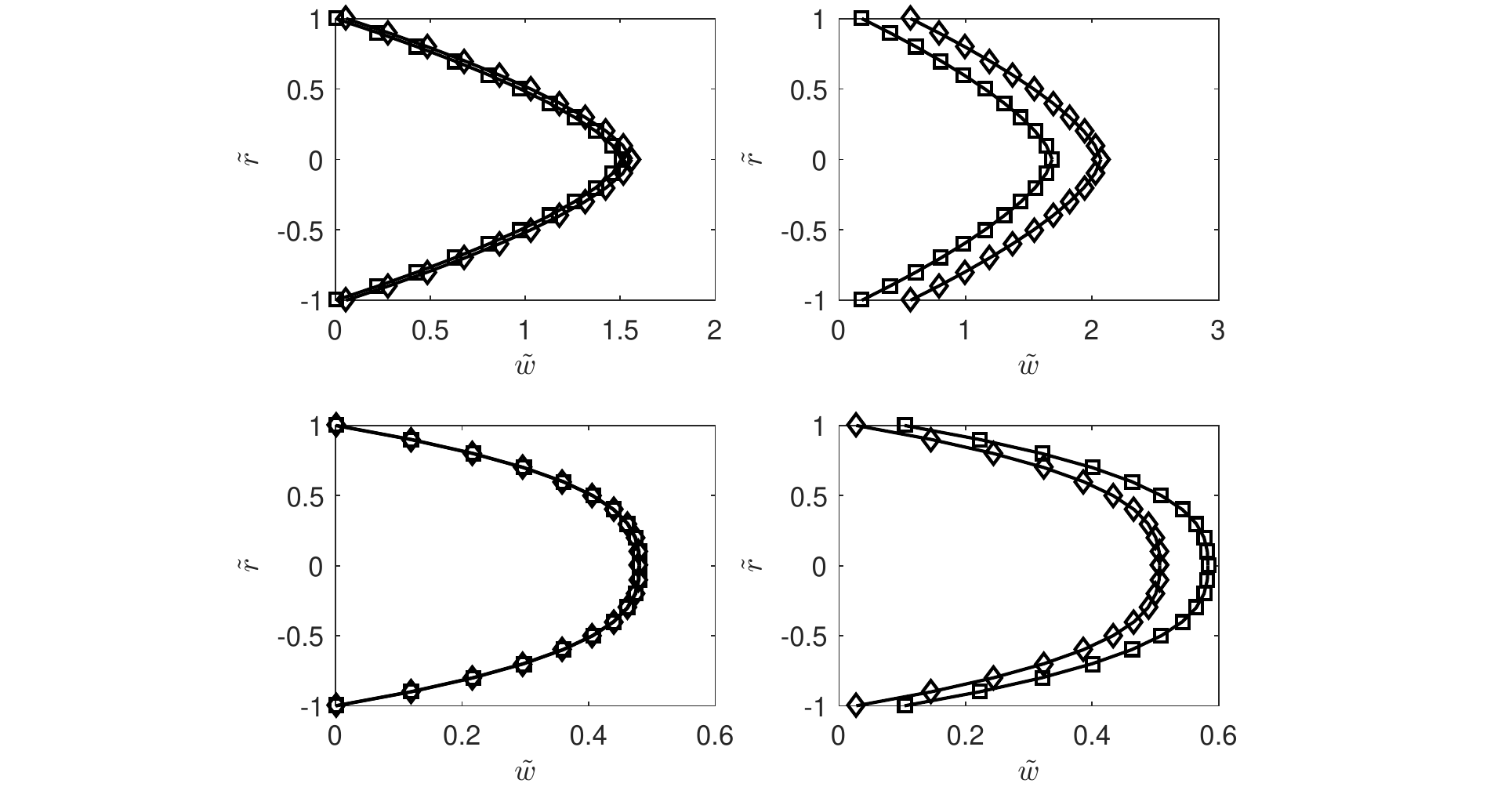}}
\caption{Non-dimensional velocities calculated using formulas (\ref{nodim_speed_NSC}) and (\ref{nodim_speed_GNSC}) for $\tilde{C}=2, \;R=0.25\, {\rm cm}$. The square symbol is used for the flow with classic Navier slip, and the diamond stands for the flow satisfying the generalized Navier slip condition. The top row corresponds to $\alpha=0,5$, while the bottom row corresponds to $\alpha=1.7$. The left column is for $l=0.0002 \,{\rm cm}$, and the right column is for $l=0.02\,{\rm cm}$.}
\label{fig8}
\end{figure}

\begin{figure}[htbp]
\centerline{\includegraphics[clip, trim=1cm 8.5cm 1cm 9cm, width=\textwidth]{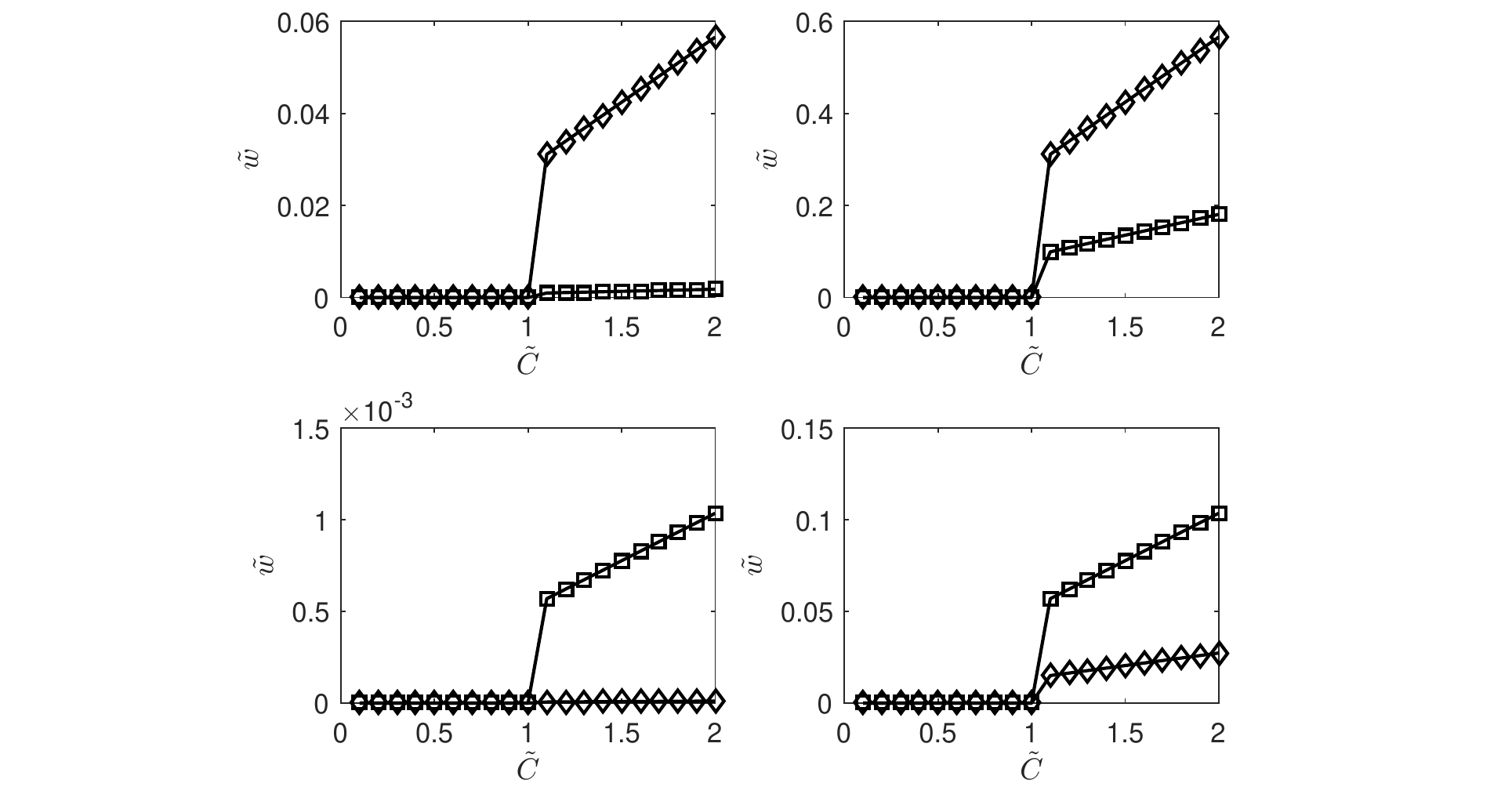}}
\caption{Non-dimensional velocities calculated at the wall using formulas (\ref{nodim_speed_NSC}) and (\ref{nodim_speed_GNSC}) for $R=0.25\, {\rm cm}$. The square symbol is used for the flow with classic Navier slip, and the diamond stands for the flow satisfying the generalized Navier slip condition. The top row corresponds to $\alpha=0,5$, while the bottom row corresponds to $\alpha=1.7$. The left column is for $l=0.0002 \,{\rm cm}$, and the right column is for $l=0.02\,{\rm cm}$.}
\label{fig9}
\end{figure}

\begin{figure}[htbp]
\centerline{\includegraphics[clip, trim=1cm 8.5cm 1cm 9cm, width=\textwidth]{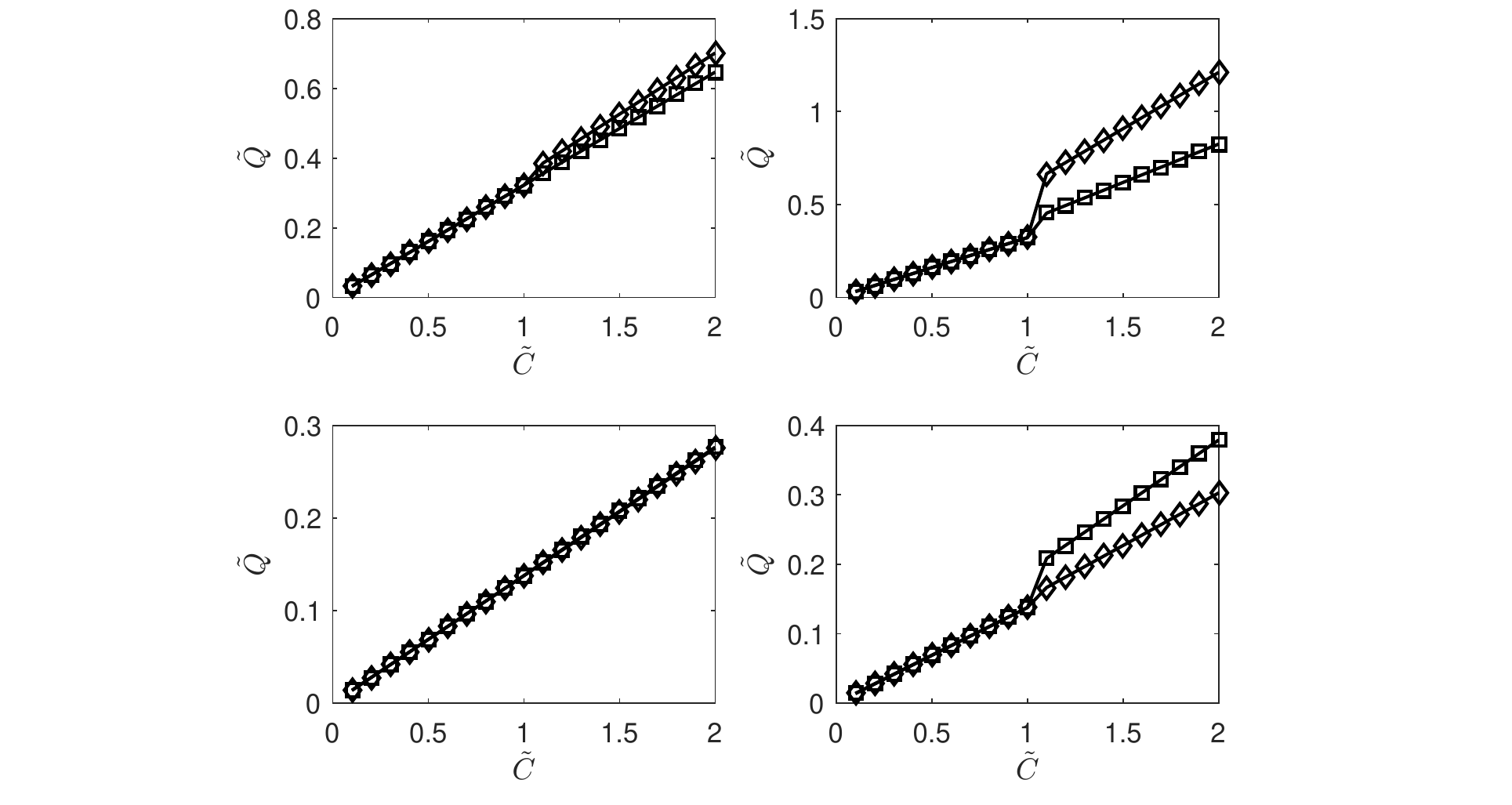}}
\caption{Non-dimensional volumetric flow rates calculated  using formulas (\ref{nodim_Q_NSC}) and (\ref{nodim_Q_GNSC}) for $R=0.25\, {\rm cm}$. The square symbol is used for the flow with classic Navier slip, and the diamond stands for the flow satisfying the generalized Navier slip condition. The top row corresponds to $\alpha=0,5$, while the bottom row corresponds to $\alpha=1.7$. The left column is for $l=0.0002 \,{\rm cm}$, and the right column is for $l=0.02\,{\rm cm}$.}
\label{fig10}
\end{figure}

Figures \ref{fig3} - \ref{fig6} represent the results obtained for $R=0.001\,{\rm cm}$, while figures \ref{fig7}-\ref{fig10} correspond to $R=0.25\,{\rm cm}$. The left column of figure \ref{fig3} shows dimensionless velocity profiles for $l=0.0002\,{\rm cm}$ for two values of $\alpha$, 0.5 and 1.7. More slip is observed in the velocity profiles satisfying the generalized Navier slip condition (bottom plot, left column of figure \ref{fig3}) than in the velocity profiles corresponding to the classic Navier slip condition (top plot, left column of figure \ref{fig3}). The first column of figure \ref{fig4} shows that for $l=0.0002\,{\rm cm}$ the velocity profiles for the two slip conditions are close for $\alpha=0.5$ (top plot, left column of figure \ref{fig4}) and $\alpha=1.7$ (bottom plot, left column of figure \ref{fig4}). However, for the characterisitc slip length $l=0.02\,{\rm cm}$ big slips are observed for both slip conditions (right columns of figures \ref{fig3} and \ref{fig4}). Not only that there are big differences between the velocity profiles for $\alpha=0.5$ and $\alpha=1.7$ for the Navier slip condition (top plot, right column of figure \ref{fig3}) and for the generalized Navier condition (bottom plot, right column of figure \ref{fig3}), but also there are big differences between the velocity profiles corresponding to two different slip conditions and the same value of $\alpha$. For $\alpha=0.5$ the velocity profile satisfying the classic Navier slip condition is about 4 times bigger than the velocity corresponding to the generalized Navier slip condition (top plot, right column of figure \ref{fig4}), while for $\alpha=1.7$ the opposite is observed but with a much larger difference between the profiles (bottom plot, right column of figure \ref{fig4}). Similar differences between the classic Navier slip and generalized Navier slip conditions are observed in the velocity profiles at the pipe wall (figure \ref{fig5}) and the corresponding dimensionless volumetric flow rates (figure \ref{fig6}) with varying $\tilde{C}$. 

When $R=0.25\,{\rm cm}$ and $\tilde{C}=2$, the velocities corresponding to the classic Navier slip and to the generalized Navier slip conditions increase with decreasing $\alpha$ for $l=0.0002\,{\rm cm}$ and for $l=0.02\,{\rm cm}$ (figure \ref{fig7}), and more slip is observed for the velocity profile satisfying the generalized Navier slip condition when $l=0.02\, {\rm cm}$ (bottom plot, right column of figure \ref{fig7}). In fact, the difference between the velocity profiles for the two slip conditions is either very close to zero when $l=0.0002\,{\rm cm}$ (left column of figure \ref{fig8}), or small when $l=0.02\,{\rm cm}$ (right column of figure \ref{fig8}). Lastly, figures \ref{fig9} and \ref{fig10} show that as $\tilde{C}$ varies, the difference between the velocity profiles at the wall corresponding to the two slip conditions is very small, and this is also valid for the associated volumetric flow rates.

\newpage

\begin{figure}[htbp]
\centerline{\includegraphics[width=\textwidth]{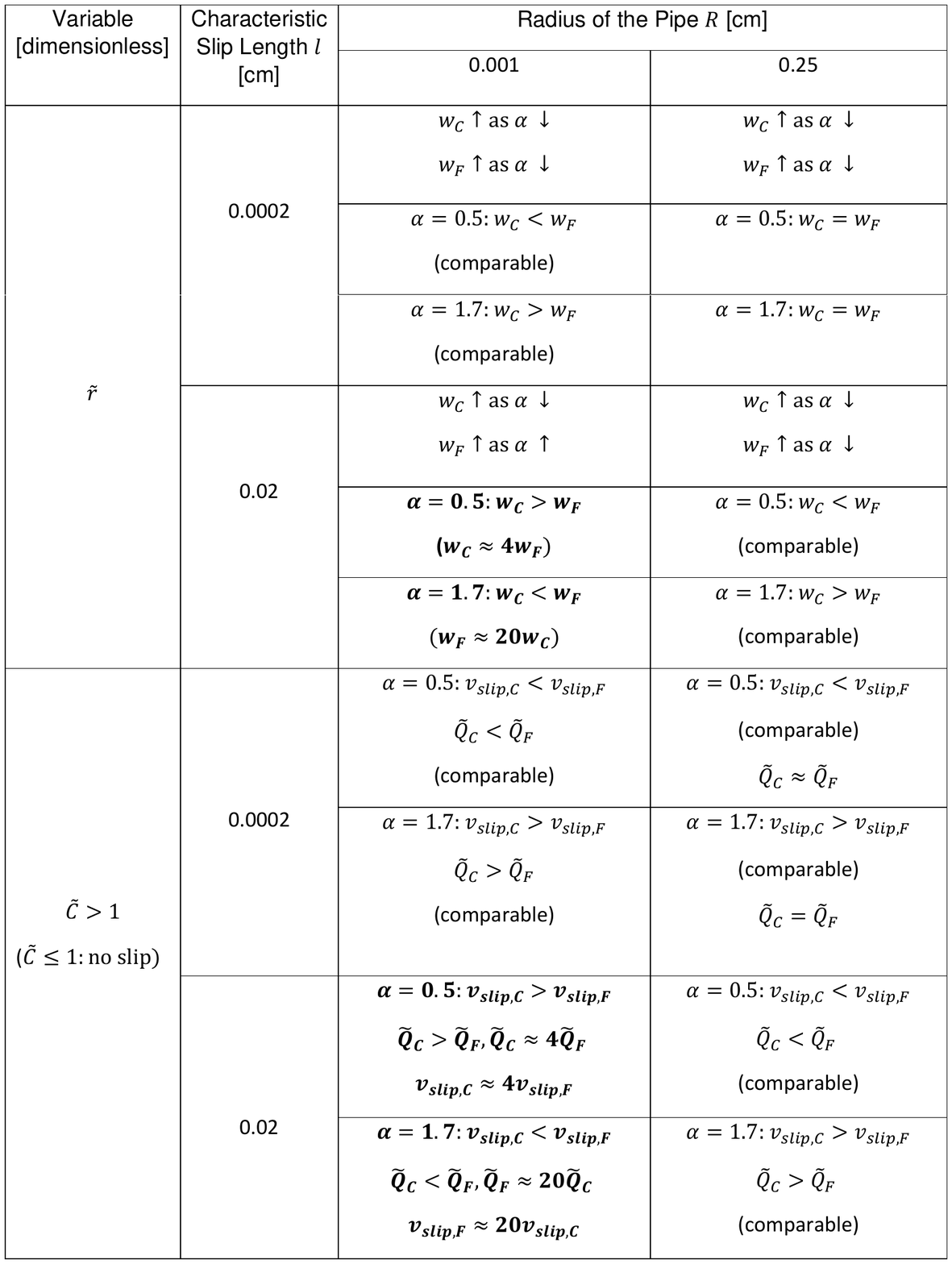}}
\caption{Review of the results. Subscript `C' stands for the classic Navier slip condition, while subscript `F' represents the results corresponding to the generalized Navier slip condition.}
\label{fig11}
\end{figure}

In figure \ref{fig11} we summarize our results presented above. Significant differences between the classic Navier slip and the generalized Navier slip conditions are observed only for $R=0.001\,{\rm cm}$ and $l=0.02\,{\rm cm}$ (bold letters in figure \ref{fig11}). The results show that a moderate increase of the dimensionless pressure gradient $\tilde{C}$ causes a dramatic increase in the dimensionless volumetric flow rate $\tilde{Q}$ which suggests that a persistent blood pressure above the critical shear stress when slips starts which could represent the presence of hypertension (by itself or due to diabetes) might contribute to the formation of microaneurysms since this big increase in $\tilde{Q}$ as $\tilde{C}$ increases could physically be achieved only by the dilatation of the arterioles. The validity of this statement will be investigated in our further work when we will consider a pipe with deformable walls. We also notice that both slip conditions could cause big volumetric flow rates but for different values of $\alpha$. It is possible that variations of $\alpha$ during flow could change the slip condition at the wall. Since we envision $\alpha$ as a measure of entanglement not only among blood cells but also among blood and endothelial cells, $\alpha$ could be changed by chemo-mechanical signaling between glial cells and these entangled arterial cells. We plan to explore this aspect in the near future.  
 
\section{Conclusions}

In this paper we proposed a mechanism for the formation, growth and rupture of microaneurysms that involve the coupled dynamics of blood, arterioles and neuroglia. The slip at the blood-vascular wall interface is seen as the initiator of this chemo-mechanical mechanism. A novel non-local constitutive law for the blood modeled as an incompressible non-Newtonian fluid is proposed and the Poiseuille flow of such a fluid through an axi-symmetric circular pipe with rigid and impermeable walls is solved analytically. Two slip conditions are considered: the classical Navier slip and a new generalized slip condition. Both, the constitutive law and the generalized slip conditon, use spatial fractional order Caputo derivatives to express aggregation and entanglement of either only blood cells or blood and endothelial cells. Our numerical simulations suggest that hypertension could contribute to the formation of microaneurysms which agrees with some clinical observations. In our future work we plan to incorporate the deformability of the pipe due to blood flow and neuroglial chemical signaling. 

\vspace{6pt} 




\end{document}